\documentclass[10pt]{article}
\usepackage[OE]{express}
\usepackage{bm}

\newcommand \avg[1][usedefault, addprefix=\global, 1=]{\langle#1\rangle}

\begin{document}

\title{Correlation steering in the angularly multimode Raman atomic memory}

\author{Mateusz Mazelanik, Micha\l{} D\k{a}browski,\authormark{*} and Wojciech Wasilewski}

\address{Institute of Experimental Physics, Faculty of Physics, University of Warsaw, Pasteura 5, 02-093 Warsaw, Poland}

\email{\authormark{*}mdabrowski@fuw.edu.pl} 

\begin{abstract}
We present the possibility of steering the direction of correlations
between the off-resonant Raman scattered photons from the angularly multimode
atomic memory based on warm rubidium vapors. Using acousto-optic
deflectors (AOD) driven by different modulation frequencies we experimentally
change the angle of incidence of the laser beams on the atomic ensemble. Performing correlations measurements for
various deflection angles we verify that we can choose the anti-Stokes light propagation direction independently of the correlated Stokes scattered light in the continuous
way. As a result we can select the spatial
mode of photons retrieved from the memory, which
may be important for future development of quantum information
processing. 
\end{abstract}



\section{Introduction}

\subsection{State of the art and motivation}

Single photon sources are essential elements of quantum information
processing (QIP) architecture. They play inevitable role in quantum
key distribution \cite{Bennett1992,Gisin2002}, quantum linear computing
\cite{Knill2001,Kok2007} and communication \cite{Duan2001}. So far
spontaneous parametric down-conversion (SPDC) sources remain most
widespread \cite{Kwiat1995,URen2004,URen2007}.

The quantum state of a photon pair produced in SPDC is well defined and quite easy
to engineer with pulsed pumping \cite{Kwiat1995}. This enables building
heralded photon sources with proper spatio-temporal properties. However,
such sources produce photons at random and rarely \cite{Broome2011},
if contribution from multiphoton states is to be negligible. Many
heralded sources can be used with a fast single-photon switch \cite{Hall2011,Ma2011}
which routes the photon from the source that luckily worked to a single
output, increasing chance of photon generation. In practice, setups
with at most four SPDC sources have been demonstrated \cite{Yao2012,Collins2013}.

Another option consists in using many SPDC sources and quantum memories
to accumulate photons and release them on demand \cite{Nunn2013}.
Once a photon is emitted in any of $M$ memory mode, one can use
an active optical switch \cite{Hall2011,Ma2011} controlled by the triggering
signal from the Stokes scattering, to route the anti-Stokes photon
to a desired single output. Such operation essentially relies on a
memory storage time which has to exceed nanoseconds-long reconfiguration
time of the optical switch \cite{Hall2011,Ma2011}. The experimental
realization of that proposal uses an external switch with a dozen distinct
atomic ensembles inside a cold quantum memory system \cite{Lan2009}.
This may be potentially useful to provide enhancement of the entanglement
swapping rate between the nodes of a quantum network \cite{Collins2007}.

\subsection{The photon source}

Here we propose a scheme equivalent to $M$ pair-sources with a built-in switch and verify its main aspects in a macroscopic light regime, with $10^3$ photons in each spatial mode.
The key element of the scheme is an angularly-multimode atomic memory \cite{Surmacz2008},
based on Raman scattering in warm atomic vapors, in which many independent
modes are stored using only one atomic ensemble  \cite{Higginbottom2012,Shuker2008}. By redirecting the
readout laser beam we change the emission angle of photons at the readout stage of the memory. 
This way we combine an equivalent multiple photon
sources, memories and a switch in a single cell. Such solution enables
redirecting the retrieved photon to a single output fiber. Note that
we operate on macroscopic light, instead of manipulating fragile quantum entities (e.g. single atomic excitations), that are prone to losses and decoherence during such a manipulation.

\begin{figure}[b]
\includegraphics[scale=0.7]{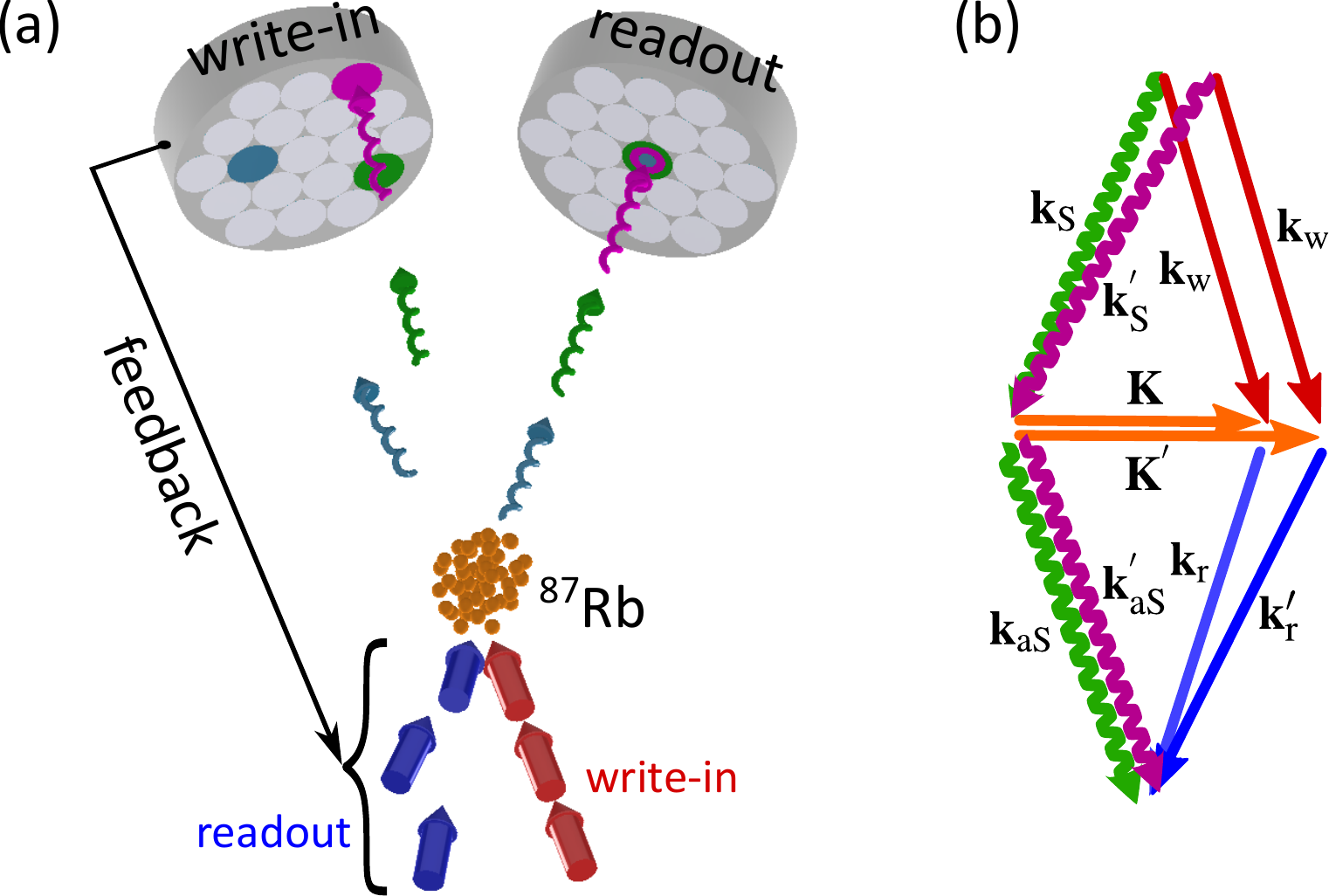}\centering

\caption{Redirecting heralded photons into the same mode in warm rubidium vapors
atomic memory. (a) Photons are scattered at random directions during
the write-in process. By redirecting readout beam we can launch the
readout photon into the same fiber each time. \mbox{(b) Phase-matching} in
the Raman scattering process. The momenta difference between the write
laser $\mathrm{\mathbf{k}}_{\mathrm{w}}$ and scattered photon $\mathrm{\mathbf{k}}_{\mathrm{S}}$
is stored in the spinwave wavevector $\mathbf{K}$. Then, it is added
to the readout laser $\mathrm{\mathbf{k}}_{\mathrm{r}}$ leading to a
well defined readout photon momentum $\mathrm{\mathbf{k}}_{\mathrm{aS}}$.
For any write-in photon direction $\mathrm{\mathbf{k}}_{\mathrm{S}}$
or $\mathrm{\mathbf{k}}_{\mathrm{S}}^{\mathrm{'}}$ and any spinwaves
with wavevectors $\mathbf{K}$ or $\mathbf{K}^{'}$, the direction
of the readout beam can be adjusted accordingly from $\mathrm{\mathbf{k}}_{\mathrm{r}}$
to $\mathrm{\mathbf{k}'_{r}}$ so as to preserve $\mathrm{\mathbf{k}}_{\mathrm{aS}}=\mathrm{\mathbf{k}}_{\mathrm{aS}}^{\mathrm{'}}$. \label{fig:Idea-of-feedback}}
\end{figure}

The concept is illustrated in Fig. \ref{fig:Idea-of-feedback}(a)
were we present three sequential realizations of the generation of a single
photon using Raman scattering in an ensemble of rubidium-87 atoms.
In each single realization we use write-in laser with the same wavevector
$\mathrm{\mathbf{k}_{w}}$ to induce Stokes scattering. The probability
of generating Stokes photon in a single spatial mode is kept very
low (typically $p\ll1/100$) but the number of modes $M$ can be high, so the probability of generating a photon in any mode $Mp$ is significant.
In some repetitions Stokes photon with wavevector $\mathrm{\mathbf{k}}_{\mathrm{S}}$
is emitted in random direction and at the same time collective atomic
excitation with wavevector $\mathrm{\mathbf{K}=\mathbf{k}_{r}-\mathbf{k}_{S}}$
is created. The atomic excitation can be stored for as long as the memory
lifetime. Finally, the readout laser converts the atomic excitation to
an anti-Stokes photon. 

The direction of the anti-Stokes photon is determined by the directions
of the Stokes photon and driving laser beams due to the time-delayed
phase-matching conditions \cite{Parniak2016,Turnbull2013} depicted in Fig. \ref{fig:Idea-of-feedback}(b).
As a result the anti-Stokes photon has a wavevector:
\begin{equation}
\mathrm{\mathbf{k}_{aS}=\mathbf{K}+\mathbf{k}_{r}=\mathbf{k}_{w}-\mathbf{k}_{S}+\mathbf{k}_{r}},\label{eq:k_as}
\end{equation}
where $\mathrm{\mathbf{k}_{r}}$ is wavevector of the readout beam. Hence, even though the direction of the Stokes photon is random, the
direction of the anti-Stokes photon can be controlled by sending the readout
beam at a proper angle. As soon as we register a Stokes photon, we
can calculate the wavevector difference $\mathbf{K}$ stored inside
the atoms. Next we compute the proper $\mathrm{\mathbf{k}}_{\mathrm{r}}$
from Eq. \ref{eq:k_as} and shine the readout laser at an angle corresponding
to this $\mathrm{\mathbf{k}}_{\mathrm{r}}$. This way the readout
photon can be launched into the same fiber each time as depicted in
Fig. \ref{fig:Idea-of-feedback}(a).

\section{Experiment}

\subsection{Off-resonant Raman scattering in $\Lambda$-system}

\begin{figure}[b]
\includegraphics[scale=0.45]{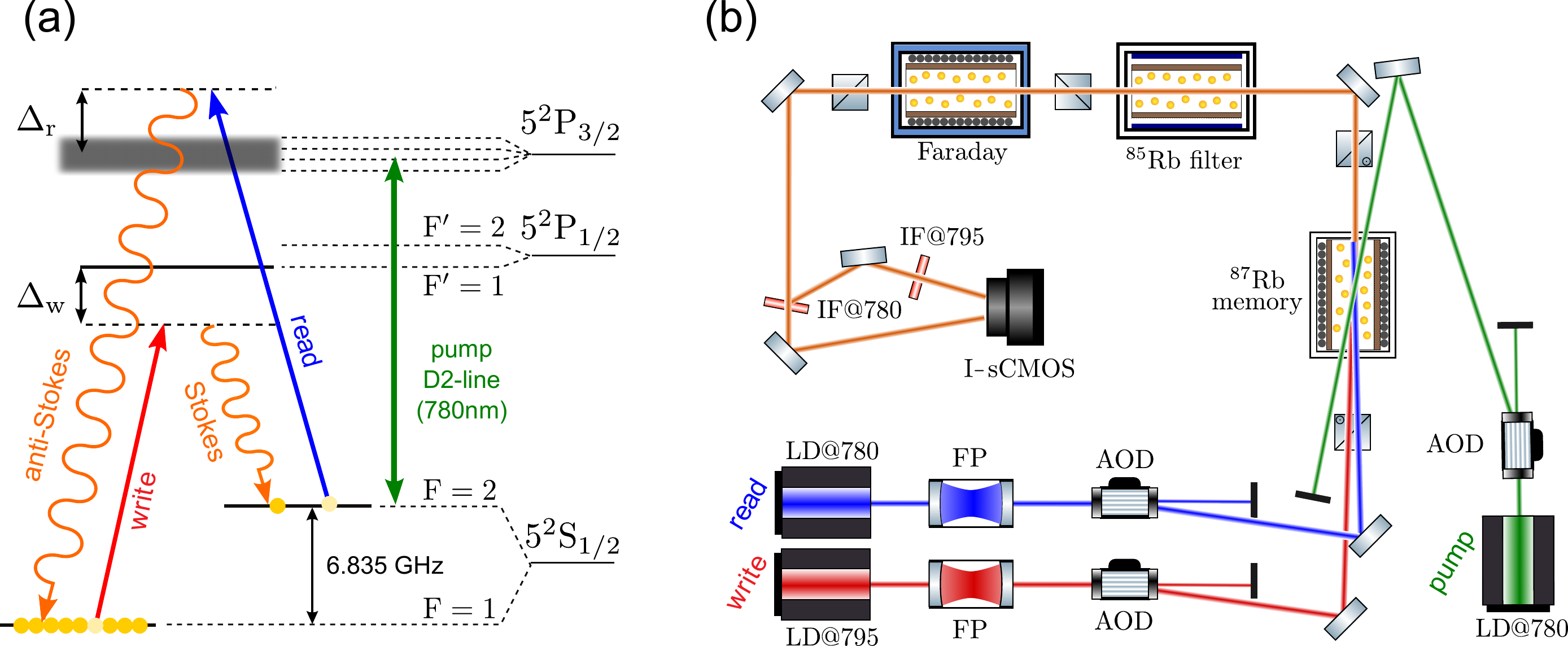}\centering\caption{(a) Atomic levels structure of rubidium-87 used in the experiment. (b)
Experimental setup: LD -- laser diodes, FP -- Fabry-P{\'e}rot interferometer,
AOD -- acousto-optic deflector, $\mathrm{^{87}Rb\ memory}$ -- atomic
memory cell, $\mathrm{^{87}Rb\ filter}$ -- absorption filter, Faraday -- Faraday filter, IF --
interference filter, I-sCMOS -- sCMOS camera with image intersifier.\label{fig:setup}}
\end{figure}

Our atomic memory scheme is based on off-resonant Raman scattering
in $\Lambda$-system of \mbox{rubidium-87} presented in the Fig. \ref{fig:setup}(a).
Atoms are initially prepared in $\mathrm{F=1}$ ground state sublevel
by the optical pumping. The pumping laser is resonant with $\mathrm{5^{2}S_{1/2},F=2\rightarrow5^{2}P_{3/2}}$
transition of the D2-line ($\mathrm{780\ nm}$). The write-in laser at
$\mathrm{\mathrm{795\ n}m}$ is detuned from the $\mathrm{F=1\rightarrow F'=1}$
transition of the D1-line to the red by $\Delta\mathrm{_{w}=1\ GHz}$
and induces Stokes scattering. In this process the Stokes photons are
created together with spinwaves corresponding to a coherent excitation
of atoms into $\mathrm{F=2}$ ground-state sublevel. We stop the write-in
process after a few microseconds during which macroscopic amount of
light is scattered. After an adjustable amount of time we illuminate
the atomic ensemble with the readout laser, which drives the anti-Stokes scattering.
This readout laser at $\mathrm{780\ nm}$ is detuned from the $\mathrm{5^{2}S_{1/2},F=2\rightarrow5^{2}P_{3/2}}$
to the blue by $\Delta\mathrm{_{r}=1\ GHz}$ and creates an anti-Stokes
photon at the expense of annihilating the atomic spinwave.

\subsection{Experimental setup details}

The experimental setup is depicted in Fig. \ref{fig:setup}(b). We
use $\mathrm{16\ mW}$ write-in and readout laser power each and $\mathrm{70\ mW}$
power of pumping laser. The $2w_{0}$ beams diameters inside memory
cell were $\mathrm{7\ mm}$, $\mathrm{7\ mm}$ and $\mathrm{12\ mm}$,
respectively. All of the lasers are frequency stabilized using DAVLL (Dichroic Atomic Vapor Laser Lock)
scheme or saturation spectroscopy. The laser beams are initially filtered
using Fabry-P{\'e}rot interferometers and then they pass through the AODs.
We use an atomic memory cell with a length of $\mathrm{L=10}$ cm (which corresponds to Fresnel number $\mathcal{F}\approx150$), containing warm $\mathrm{^{87}Rb}$
vapors with 1 Torr pressure krypton as a buffer gas \cite{Chrapkiewicz2012},
heated to about $\mathrm{80\ ^{\circ}C}$. The output light of the
atomic memory cell is filtered out using a Wollaston prism, atomic
absorption filter and a Faraday filter \cite{Dabrowski2015}. The
temperature of atomic absorption filter is set to $\mathrm{100\ ^{\circ}C}$.
Faraday filter also has a temperature of $\mathrm{T=100\ ^{\circ}C}$
and magnetic field set to $\mathrm{B=12\ \mathrm{mT}}$. Both the
atomic memory cell and the Faraday filter were inside two layer magnetic
shielding. The detunings ($\Delta\mathrm{_{w}},\Delta\mathrm{_{r}})$, the Faraday
filter temperature and magnetic field was set to obtain the best signal
to noise ratio. The Stokes was separated from the anti-Stokes light using
$\mathrm{795\ nm}$ and $\mathrm{780\ nm}$ interference filters and both were
directed into two separate regions on the camera. The camera pane could be divided into separate circular areas corresponding to distinct virtual fibers, presented in Fig. \ref{fig:corr_maps}.  The pulse sequence
consisted of $\mathrm{\tau_{p}=350\ \mu s}$ pump pulse and $\mathrm{\tau_{w}=\tau_{r}=8\ \mu s}$
write-in and readout pulses, all of a rectangular shape. Memory storage
time, i.e. delay between write-in and readout was equal to $\mathrm{\tau_{s}=1\ \mu s}$.
During the pulse sequence the cell heating was turned off to avoid
the magnetic fields coming from the heating coils. Finally, the scattered
photons were detected with an sCMOS (scientific Complementary Metal-Oxide Semiconductor) camera equipped with an image intensifier (I-sCMOS) \cite{Chrapkiewicz2014a,Jachura2015}. The gate of the image intensifier
 was open for the combined duration of the write-in,
storage and readout pulses \cite{Dabrowski2014}. The image intensifier operated in the
linear response regime.

\subsection{Acousto-optic deflector (AOD)}

\begin{figure}[b]
\includegraphics[scale=0.45]{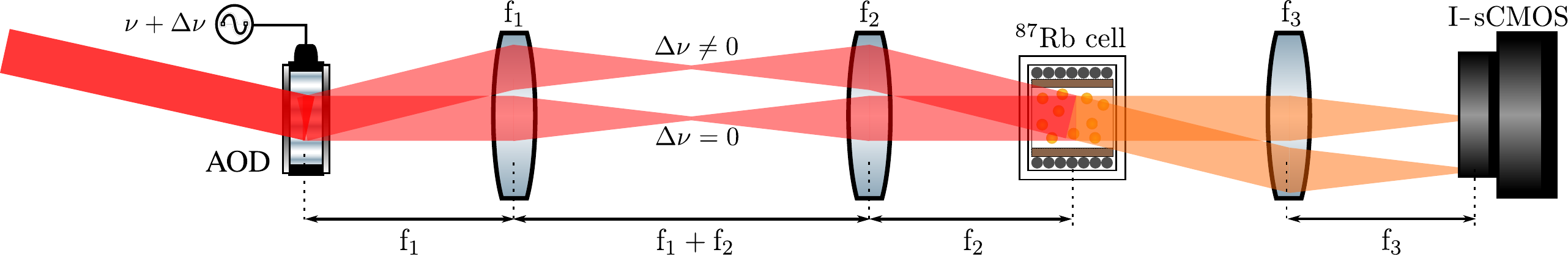}\centering\caption{The optical 4f relay transfers the angle of deflection of AOD into
the centre of atomic memory. The deflection angle is proportional
to AOD electrical drive frequency. The centre of AOD is optically
imaged onto the center of atomic memory cell and next the crossing
beams are projected onto I-sCMOS camera pane situated in the far-field
with respect to the memory cell. The focal length of lenses: $\mathrm{f_{1}=50\ mm}$,
$\mathrm{f_{2}=750\ mm}$, $\mathrm{f_{3}=500\ mm}$. Basic modulation
frequency was set to $\mathrm{\nu=80\ MHz}$, the laser beam diameter
at the AOD input $w=\mathrm{250\ \mu m}$. Such a configuration does
not change the position of beams intersection inside memory cell but
only changes the angle of crossing thus effectively changing the beams
position in the far field camera. \label{fig:The-idea-of-AOM-1}}
\end{figure}

Acousto-optical deflectors (AOD) were used to steer the laser beams
directions \cite{Roberts2014} and produce microsecond driving pulses.
The deflection angle is transferred into the center of the memory cell by
a 4f relay lens system presented in Fig. \ref{fig:The-idea-of-AOM-1}.
This particular deflection angle is changing by reconfiguring the electronic
direct digital synthesizer (DDS), driving the deflector at a desired frequency.
At last, the $\mathrm{f_{3}}$ lens focuses scattered photons, converting scattering
angle into a position on the I-sCMOS sensor.

One can notice that in present solution beams deflected at different angles have different frequency.
This frequency difference applies to anti-Stokes photons retrieved from the memory.
However, this effect can be compensated by using  additional  AOD set in double pass system.

\section{Results}

\subsection{Correlation measurements using camera}

To verify the direction control ability of our photon source we calculated
the correlations between the directions of scattered photons. In each
iteration of the experiment we register an image formed by thousands
of scattered photons which corresponds to an intensity of light $\mathrm{I(\mathrm{\bm{\theta})}}$
emitted at a certain direction $\mathrm{\bm{\theta}=(\theta_{x},\theta_{y})}$.
In the following we plot maps of correlation coefficient $\mathrm{C(\bm{\theta},\bm{\theta}^{'})}$
between light emitted at a fixed angle $\mathrm{\bm{\theta}^{'}}$and
light emitted at any other angle $\mathrm{\bm{\theta}}$  \cite{Dabrowski2014,Chrapkiewicz2012}:

\begin{equation}
\mathrm{C(\bm{\theta},\bm{\theta}^{'})}=\frac{\avg[\mathrm{I}(\bm{\theta})\mathrm{I}(\bm{\theta}^{'})]-\avg[\mathrm{I}(\bm{\theta})]\avg[\mathrm{I}(\bm{\theta}^{'})]}{\sqrt{\avg[(\Delta \mathrm{I}(\bm{\theta}))^{2}]\avg[(\Delta \mathrm{I}(\bm{\theta}^{'}))^{2}]}},\label{eq:corr}
\end{equation}
where: $\mathrm{I(\bm{\theta})}$ is the mean intensity of light
registered at angle $\bm{\theta}$ and $\Delta\mathrm{I}(\bm{\theta})$
is deviation from this mean value.
The angles $\mathrm{\bm{\theta}}$ are related to $\mathrm{x}$
and $\mathrm{y}$ components of wavevectors in Eq.~(\ref{eq:k_as})
via a simple relation valid for small angles $\mathrm{\mathrm{2\pi\bm{\theta}=}\mathbf{k}_{\perp}\lambda}$,
where $\mathrm{\lambda}$ is the wavelength of the driving laser beam.

In the low gain regime one would measure the number of coincidences $n(\bm{\theta},\bm{\theta^{'}})$ between the Stokes photon emitted at the angle $\bm{\theta^{'}}$ and the anti-Stokes photon at the angle $\bm{\theta}$ \cite{Chrapkiewicz2016}. In high gain regime pairing up photons is no longer possible, and a number of photons emitted in any direction appears to be random. However, the number of photons (i.e. intensity of light) emitted at angle $\bm{\theta}$ in each realization of the experiment should be in the ideal case equal to the number of photons emitted in the conjugate direction  $\bm{\theta^{'}}=-\bm{\theta}$.

\subsection{Correlation maps}

\begin{figure}[h!]
\centering\includegraphics[scale=0.36]{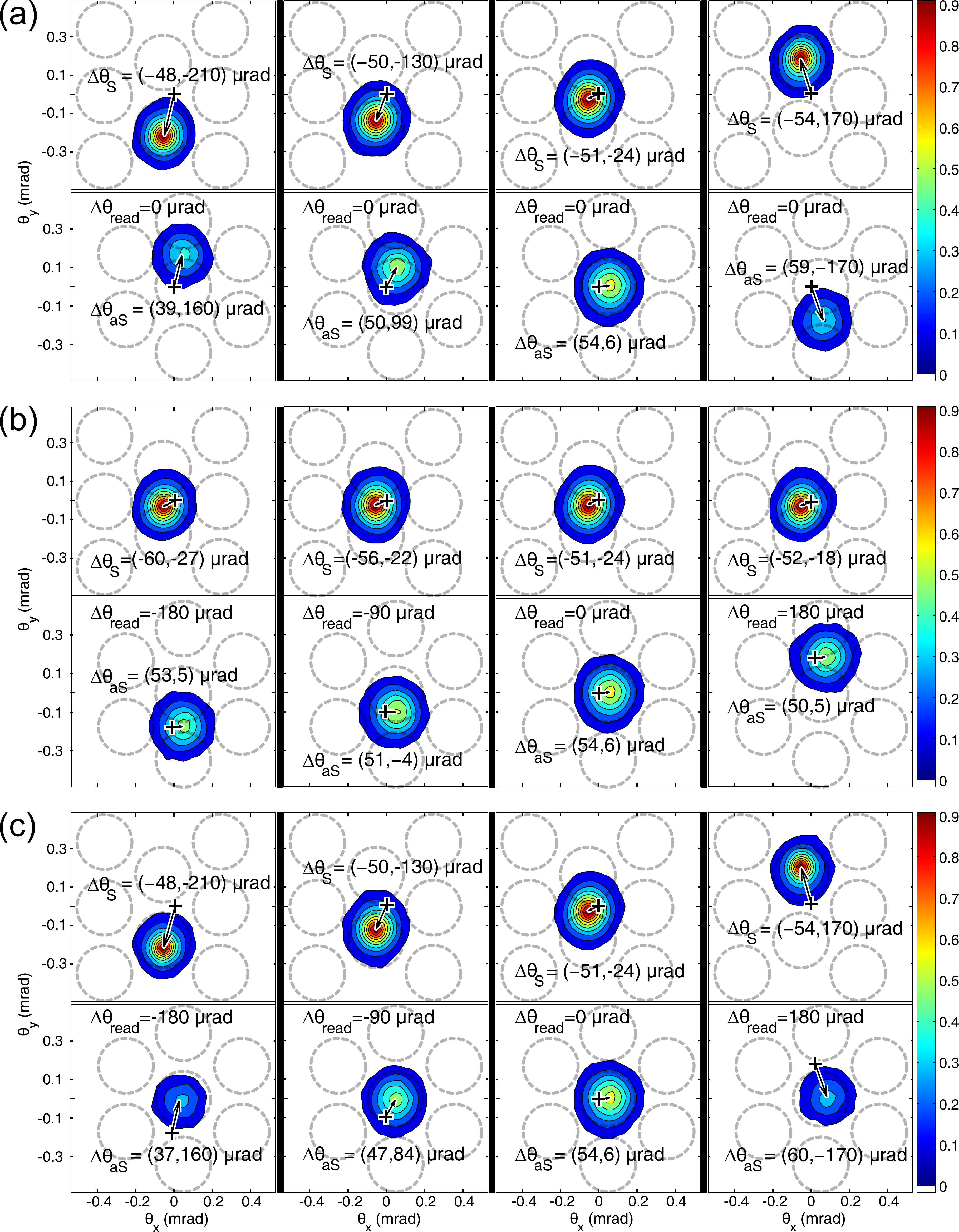}

\caption{Maps of correlations between intensity registered at a reference angle
and the registered far field images around Stokes (write-in) driving
beams (upper parts) and anti-Stokes (readout, lower parts). Rows represent
situations with: (a) different reference angles $\Delta\mathrm{\mathrm{\bm{\theta}_{S}}}$,
(b) different AOD deflection angles $\Delta\mathrm{\theta_{read}}$,
(c) combination of reference points from (a) and deflection angles
from (b) simultaneously, leading to static position of the twin spot center.
That coresponds to coupling the light always into the same fiber of conceptual fiber matrix, depicted as dashed circles.\label{fig:corr_maps}}
\end{figure}

We measured the scattering angles for the direction of the write-in laser
which was fixed during the experiment and central direction of the
readout laser. When deflected, the readout laser was sent at angle
$\mathrm{\bm{\theta}}_\mathrm{read}=(0,\Delta\mathrm{\theta_{read})}$, where $\Delta\mathrm{\theta_{read}}$
was controlled by the AOD system. In Fig. \ref{fig:corr_maps}
we present maps of intensity correlation coefficient $\mathrm{C(\bm{\theta},\bm{\theta}^{'})}$
calculated from $10^{4}$ camera frames.

Upper and lower parts of
each panel depict correlations with light scattered during write-in
and readout, respectively. Directions of the write-in and readout laser beams are marked as black crosses. The reference point $\mathrm{\bm{\theta}^{'}}=\Delta\bm{\theta}_{\mathrm{S}}$
is chosen in the write-in part of the camera pane. It can be identified with the position of the virtual fiber depicted as dashed circle that collects the heralding photons. The first row in the Fig. \ref{fig:corr_maps}(a) presents correlation
maps for stationary drive beams. In each panel we take different reference
point $\mathrm{\bm{\theta}^{'}}=\Delta\bm{\theta}_{\mathrm{S}}$.

Registration of a photon at a reference point corresponds to projecting
the state of the atoms onto a spinwave. The spinwave is clipped to
the size of the write-in beam of diameter $2w_0$ and therefore its
wavevector $\mathbf{K}$ is defined with a finite precision $\mathrm{\pm\lambda}/w_0$.
This spinwave stimulates scattering of subsequent photons into a finite
solid angle, corresponding to a certain angular spread of photons.
The stimulated photons give rise to a finite-sized spot of Stokes-Stokes
correlations around the reference direction $\Delta\mathrm{\bm{\theta}_{S}}$
in upper parts of each panel. Note that they move around the reference
direction. The twin spot located in the lower part of each panel attests
the presence of correlated anti-Stokes photons. It moves in opposite
direction $\Delta\mathrm{\bm{\theta}_{aS}\approx-}\Delta\mathrm{\bm{\theta}_{S}}$,
according to the phase-matching condition described by Eq. (\ref{eq:k_as}).
Due to finite spatial size of the spinwave the anti-Stokes photons
are emitted into a finite solid angle. Their angular spread is virtually
the same as for the Stokes photons because they come from the same
spinwave.

With Fig. \ref{fig:corr_maps}(a) we verify that by registering a
Stokes photon the direction of the anti-Stokes photon can be predicted.
We can also see finite angular spread of the correlated photons which
is a signature of a mode size and should match fiber mode to obtain
best heralding efficiency. The number of modes M can be approximated as twice the number of correlation spots that fit into the solid angle occupied by the scattered light  \cite{Chrapkiewicz2012}. In the experiment we obtain $M=20$ and $M=10$ for write-in and readout processes, respectively. Virtual fibers depicted in Fig. \ref{fig:corr_maps} are placed with empty spaces between each other, thus the number of fibers is effectively lower than the number of modes.

In the second row of Fig. \ref{fig:corr_maps}(b)
we demonstrate manipulation of the emission angle of anti-Stokes light by
changing the direction of readout laser $\Delta\mathrm{\theta_{read}}$.
We keep the reference point $\mathrm{\bm{\theta}^{'}}=\Delta\bm{\theta}_{\mathrm{S}}$
constant and thus the Stokes-Stokes correlations remain in place. The differences between this parameter in between panels are just the result of the uncertainty fitting Gaussian to the real data.
However, the twin spot moves together with the read beam with constant
scattering angle $\Delta\mathrm{\bm{\theta}_{aS}}$, just
as predicted by Eq. (\ref{eq:k_as}).

Finally, we verify the key aspect of the concept depicted in Fig. \ref{fig:Idea-of-feedback} using macroscopic Raman scattered light, with $10^3$ photons generated per each spatial mode. In each panel of Fig.
\ref{fig:corr_maps}(c) we change reference point corresponding to
a detection of a photon at different angle $\Delta \mathrm{\bm{\theta}_{S}}$.
By redirecting the read beam we are able to redirect the twin photons
into the same direction chosen arbitrary as $\mathrm{(54,6)\:\mu rad}$.

\subsection{Cross-sections of correlation maps}

Figure \ref{fig:Correlation-crosssection-small-1} depicts $\mathrm{\theta}_{\mathrm{y}}$
cross-sections through twin beam correlation spot from Fig. \ref{fig:corr_maps}
with fitted Gaussian curves. The angular spread of the correlation
spot is about $\mathrm{240\ \mu}$rad FWHM. This correspond to an
emission from an area of about 3.5 mm diameter. It is smaller by a
factor of 2 than the write-in beam size due to exponential character
of the Stokes scattering gain process. The spinwave is amplified primarily
in the center of the laser beam and effectively shrinks. With our AOD
system we can deflect anti-Stokes emission by about $\mathrm{400\ \mu rad}$
with precision much better than the angular spread of the correlation.
Figure \ref{fig:Correlation-crosssection-small-1}(c) confirms that
the anti-Stokes emission can be redirected into the same angular mode
regardless of the position of the trigger photon. In Fig. \ref{fig:Correlation-crosssection-small-1}(b)
the maximum correlation value drops slightly for larger deflection
angles $|\Delta\mathrm{\theta_{read}}|$. This is the result of aberrations
of our optical system. In Fig.~\ref{fig:Correlation-crosssection-small-1}(a)
the maximum value of correlation coefficient drops at higher angles
of scattering $\Delta\mathrm{\mathrm{\bm{\theta}_{S}}}$ because
of diffusional damping of spinwaves with long wavevectors \cite{Chrapkiewicz2012,Chrapkiewicz2014c}.

\begin{flushleft}
\begin{figure}[b]
\centering\includegraphics[scale=0.4]{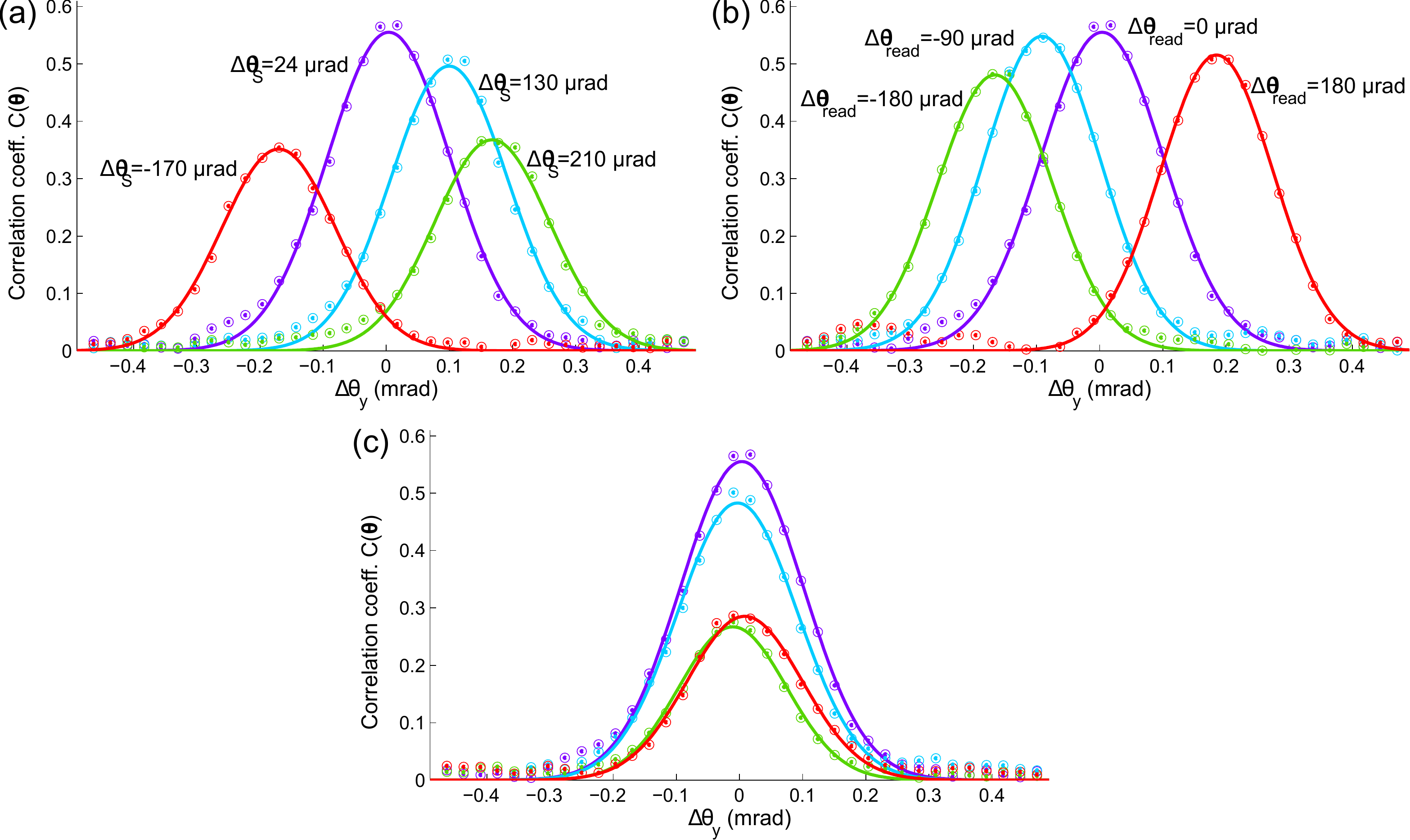}

\caption{Cross-sections of correlation maps from Fig. \ref{fig:corr_maps}(a)-(c).
The FWHM of correlation peak is about $\mathrm{0.24\ mrad}$. \label{fig:Correlation-crosssection-small-1}}
\end{figure}

\par\end{flushleft}

\subsection{Implications}

The experimental results validate the scheme proposed in Fig.~\ref{fig:Idea-of-feedback}.
The results presented in Figs. \ref{fig:corr_maps} and \ref{fig:Correlation-crosssection-small-1}
enable designing the collection optics matching the photons to the single
mode fibers. The focal length of the $\mathrm{f_{3}}$ lens should be chosen
such that the size of the correlation spot matches the size of the
fiber mode. Registering the trigger photon in any of the fibers heralds
creation of a spinwave that covers always the same volume of the cell, but
has different wavefront directions, dependent of which fiber was hit. Its spatial phase corresponds
to a plane-wave with wavevector $\mathbf{K}$. By adjusting the direction
of the readout beam we can compensate spatial phase differences. That
is between subsequent operations the wavevector of the readout beam
$\mathrm{\mathbf{k}}_{\mathrm{r}}$ is changed by precisely the same
amount by which the spinwave wavevector $\mathbf{K}$ is changed (compare Eq.~(\ref{eq:k_as})).
This way at the readout stage the anti-Stokes field is emitted from
the same volume and with the same spatial phase, always in the same
spatial mode. Figures \ref{fig:corr_maps}(c) and \ref{fig:Correlation-crosssection-small-1}(c)
effectively depict directional intensity profile of this mode. A proper
optical setup can focus the anti-Stokes light into a spot matching
always the same single mode fiber. By adjusting the write-in duration
or power, the mean number of excitations $\zeta$ created in each
twin beams can be adjusted. By lowering this number to $\zeta\lesssim 1/100$
we ensure, that in less than this amount of cases two photons would
be generated instead of one, thus conditionally creating well defined
single photon mode which state might be easily destroyed during the manipulation.

\section{Conclusions}

We proposed a scheme for a single photon source equivalent to $M$ heralded SPDC sources \cite{Nunn2013} and a switch realized in a single multimode
emisive quantum memory. Similarly as in SPDC sources, the pair production
rate per mode can be kept low enough to avoid double pair production.
At the same time the total probability for photon generation in any mode
can be high. In practice, demonstration of the whole scheme at the single photon level with real-time feedback requires better long term setup stability and faster detection time response of the sCMOS camera.

We populate random mode of the memory by driving Raman Stokes scattering.
Upon registering the direction of scattered photon the mode which
was excited is identified. Then the readout beam direction is adjusted
and this way the readout is always launched in the same direction.
The last step is an effective demultiplexing switch realized without
directly operating on the fragile single photons. 

The experimental verification demonstrated bases on measuring spatial
intensity correlations. Our 1 $\mu$s storage time
is sufficient to switch the direction of the readout beam \cite{Lan2009} and realize
the scheme. In our demonstration we can retrieve about $M=10$ modes
corresponding to independent pair sources. In the experiment, the number of excited modes is limited by the diffusion \cite{Chrapkiewicz2014c}. Therefore we expect the number of retrieved modes $M$ to be lower  than the Fresnel number $\mathcal{F}$ of the write beam. We envisage that further increasing the drive beam sizes can lead to $M=1000$ modes,
which would result in virtually deterministic single photon generation \cite{Chrapkiewicz2016}. 

\section*{Funding}
Polish National Science Centre projects no. DEC-2011/03/D/ST2/01941 and DEC-2015/19/N/ST2/01671.

\section*{Acknowledgments}

We acknowledge A. Leszczy\'{n}ski, M. Lipka and M. Parniak for discussions about the manuscript. 
\end{document}